\newcommand{\bee}{\begin{equation}}
\newcommand{\ene}{\end{equation}}
\newcommand{\beea}{\begin{eqnarray}}
\newcommand{\enea}{\end{eqnarray}}

\documentclass[aps,preprint,pre,showpacs,superscriptaddress]{revtex4}
\usepackage{amsmath} 
\usepackage{graphicx}
\usepackage{dcolumn}
\usepackage{bm}
\begin{document}
\title{\textbf{Collective Free Electron Excitations in Half-Space Configuration}}
\author{M. Akbari-Moghanjoughi}
\affiliation{Faculty of Sciences, Department of Physics, Azarbaijan Shahid Madani University, 51745-406 Tabriz, Iran}

\begin{abstract}
In this research, using the suitably linearized kinetic-corrected Schr\"{o}dinger-Poisson system, we study the first-order density and potential perturbations in a half-space electron gas of arbitrary degeneracy in the jellium model. By solving the eigenvalue problem in coupled pseudoforce model with appropriate boundary conditions, pure states of plasmonic excitations in the quasineutral electron gas are determined from which mixed states of the system characterizing the thermodynamic equilibrium is constructed by enumeration of quasiparticle orbital. It is shown that the Friedel-like electron density fringes appear inside the electron gas due to interference effect. It is also remarked that screened electron packet appears in front of the physical boundary which is different from continuous density spectrum. The electrostatic potential coupled to the electron density beyond the physical boundary becomes oscillatory for density-temperature regimes relevant to fully degenerate electron gas. The potential minimum found in front of physical boundary confirms Casimir-Polder-like force between parallel metallic plates and may be responsible for agglomeration of metallic nanoparticles in quantum contacts. The dependence of potential and electron density at the electron gas boundary is investigated for a wide range of parametric density-temperature regimes. It is also found that the kinetic correction to the model has a profound effect on the particle-like low phase-speed matter-wave dispersion branch of plasmon excitations, where, the increase in the density of electrons leads to the increase in the effective electron mass, and consequently, decrease in the electron mobility, while, increase in the temperature causes an increase in the effective electron mobility.
\end{abstract}

\date{\today}
\email[Corresponding author: ]{massoud2002@yahoo.com}
\maketitle
\section{Introduction}

Plasmonics \cite{man1,maier} is a new interdisciplinary field of research with important applications in nanoelectronics \cite{haug,gardner}, optoelectronics \cite{wooten}, and semiconductor integrated circuit industry \cite{markovich,manvir}. Plasmons are elementary collective excitations of quantum plasmas which play an inevitable role in physical phenomena \cite{krall,stenf1}. These entities provide ideal platform for fast THz device communications \cite{umm} and beyond where the conventional wired communications fail to operate. Plasmonic energy conversion is an alternative way of solar power extraction in future photovoltaic and catalytic designs due to their much higher efficiency compared to semiconductor technologies \cite{cesar,jac}. The state of the art nanocatalitic engineering for energy extraction is based on conversion of energy of collective surface electron plasma oscillations caused by localized electromagnetic radiation into hot electron generation in plasmonic devices which operates in ultraviolet-visible (UV-VIS) frequency range. However, the conventional solar cell devices are based on low energy interband electron-hole transitions in semiconductors which operate at visible-near infrared (VIS-NIR) low frequency range. However, plasmonic devices operating in ultraviolet spectrum in the wavelength range ($10-380$)nm requires special technological considerations and nanofabrication costs \cite{at} due to low contribution of plasmonic device operation range to the solar radiation spectrum. Recent studies have revealed that nanostructured materials can be used to directly convert collected electromagnetic energy into electricity by local surface plasmon resonance (LSPR) in the hot electron generation mechanism \cite{at2}. The generated hot electrons are then collected by putting the metallic nanoparticles in contact with a good electron acceptor like $TiO_2$ in Schottky configuration \cite{tian}. Studies also reveal that hot electron extraction efficiency is strongly dependent on the size and geometry of nanoparticles as well as their composition \cite{vin}.

A wide range of quantum plasma regime extends from doped semiconductors \cite{hu1,seeg} with moderately high electron number density at low temperature to high density metallic compounds up to high-temperature and density so-called the warm dense matter (WDM) \cite{glenzer2,koenig}. Quantum effect arise naturally when the interspecies distances are lowered beyond the thermal de Broglie's wavelength \cite{bonitz}. In inertial confinement fusion experiment quantum effects vary in strength due to change in equation of state (EoS) of matter during shock compression causing increase in temperature and number density \cite{mithen}. The change of EoS has fundamental effect on the faith of compact stellar object setting distinct limit on the mass of such solar entities \cite{chandrasekhar1}. Another interesting feature of charged particles is their complex collective electromagnetic interactions with their average interparticle distances beyond the quantum limit. The later is due to the nonlocal nature of interactions via the Bohm's quantum potential. Such a collective interactions lead to many interesting features of electron interactions with ionic lattice in solids \cite{kit,ash}. The theoretical development of quantum plasmas has a long history with pioneering contributions over the past century \cite{bohm,pines,klimontovich}. Because of the dominant quantum effects caused by the EoS of a Fermi gas, semiclassical plasma theories which incorporate the quantum statistical pressure effects ignoring the electron diffraction can lead to some interesting collective features in dense plasmas \cite{ichimaru1}.

There has been an increasing interest in the quantum plasma research over the recent decade due to their better description of collective features in dense plasmas \cite{se,sten,ses,haas1,brod1,mark1,man3,mold2,sm,fhaas}. Quantum kinetic \cite{man2}, quantum (magneto)hydrodynamic \cite{haasbook} and gradient corrected theories \cite{bonitz0} has been used to investigate different aspects of many-fermion interactions in plasmas. Application of quantum hydrodynamic has led to discovery of many interesting collective properties of plasmas which has not been captured by theories ignoring the quantum electron diffraction feature. The quantum hydrodynamic formulations is cast into the Schr\"{o}dinger-Poisson model \cite{manfredi} using the Madelung transformation \cite{madelung}. The leter model has been used to study various linear and nonlinear features of plasmon excitations in quantum plasmas \cite{mannew,hurst}. Recently, the pseudoforce system derived from linearized Schr\"{o}dinger-Poisson model, has been used to study different features of pure states of elementary excitations in an electron gas of arbitrary degenerate electron gas \cite{akbquant}. The pseudoforce model has a fundamental property of taking into account the dual lengthscale nature of plasmons in the electron gas. The discrepancy between quantum hydrodynamic and density functional theories in the linear limit \cite{akbhd,mold,michta} has led to some kinetic correction to the fluid theories in the quantum potential term. In current analysis we take into account the kinetic correction in the Schr\"{o}dinger-Poisson model in order study the plasmonic excitations in the half-space electron gas and investigate the effects of such density-temperature dependent correction on the energy dispersion of plasmonic excitations.

\section{The Mathematical Model}

The closed set of quantum hydrodynamic equations \textbf{including the kinetic correction} to the electron quantum diffraction effect is as follows
\begin{subequations}\label{gs}
\begin{align}
&\frac{{\partial n}}{{\partial t}} + \frac{{\partial n u}}{{\partial x}} = 0,\\
&\frac{{\partial u}}{{\partial t}} + u\frac{{\partial u}}{{\partial x}} = \frac{e}{m}\frac{{\partial \phi }}{{\partial x}} - \frac{1}{m}\frac{{\partial \mu }}{{\partial x}} + {\rm{ }}\frac{{{\zeta\hbar ^2}}}{{2{m^2}}}\frac{\partial }{{\partial x}}\left( {\frac{1}{{\sqrt n }}\frac{{{\partial ^2}\sqrt n }}{{\partial {x^2}}}} \right),\\
&\frac{{{\partial ^2}\phi }}{{\partial {x^2}}} = 4\pi e\left( {n - {n_0}} \right),
\end{align}
\end{subequations}
\textbf{in which $n$, $u$, $\phi$ and $\mu$ denote the electron number density, fluid velocity, electrostatic potential and chemical potential. Here $\zeta$ represents the temperature and density dependent correction factor in order to account for the correct contribution to the low-frequency and long-wavelength plasmon excitations\cite{haas2016}.}
\begin{equation}\label{xi}
\zeta  = \frac{{{\rm{L}}{{\rm{i}}_{3/2}}\left[ {-\exp (\beta {\mu _0})} \right]{\rm{L}}{{\rm{i}}_{ - 1/2}}\left[ {-\exp (\beta {\mu _0})} \right]}}{3{{\rm{L}}{{\rm{i}}_{ 1/2}}{{\left[ {-\exp (\beta {\mu _0})} \right]}^2}}},
\end{equation}
in which $\beta=1/k_B T$ and $\mu_0$ is the equilibrium value of the chemical potential. The function ${\rm{Li}}_{k}$ is the polylog function defined below
\begin{equation}\label{pl}
{\rm{L}}{{\rm{i}}_k}[ - \exp (\eta )] =  - \frac{{{F_{k - 1}}(\eta )}}{{\Gamma (k)}},
\end{equation}
where $\Gamma$ is the gamma function and the polylog function definition in terms of the Fermi function is
\begin{equation}\label{f}
{F_{k-1}}(\eta ) = \int_0^\infty  {\frac{{{x^{k-1}dx}}}{{\exp (x - \eta ) + 1}}}.
\end{equation}
The one-dimensional Schr\"{o}dinger-Poisson model \cite{manfredi} is mathematically equivalent to the quantum hydrodynamic equations (\ref{gs}).
\begin{subequations}\label{sp}
\begin{align}
&i\sqrt{\zeta}\hbar \frac{{\partial \cal N}}{{\partial t}} =  - \zeta\frac{{{\hbar ^2}}}{{2m}}\frac{{\partial ^2 {{\cal N}}}}{{\partial {x^2}}} - e\phi{\cal N} + \mu(n,T){\cal N},\\
&\frac{{\partial ^2 {\phi}}}{{\partial {x^2}}} = 4\pi e (|{\cal N}|^2 - n_0),
\end{align}
\end{subequations}
\textbf{where ${\cal N} = \sqrt{n(x,t)}\exp[iS(x,t)/\hbar\sqrt{\zeta}]$ and $u(x,t)=(1/m)\partial S(x,t)/\partial x$ are the Madelung transformations \cite{madelung,bohm0}. Also, $n_0$ is the unperturbed electron density which is equal to that of the neutralizing positive background. It is straightforward to show the equivalence of (\ref{sp}) and (\ref{gs}) by using the ansatz transformations. Using ${\cal N} = \sqrt{n(x,t)}\exp[iS(x,t)/\hbar\sqrt{\zeta}]$ in \ref{sp}(a) and separating real and imaginary parts, we have}
\begin{subequations}\label{sph}
\begin{align}
&m\frac{{\partial n(x,t)}}{{\partial t}}+\frac{{\partial n(x,t)}}{{\partial x}}\frac{{\partial S(x,t)}}{{\partial x}}+n(x,t)\frac{{{\partial ^2}S(x,t)}}{{\partial {x^2}}} = 0,\\
&\frac{{{\partial ^2}S(x,t)}}{{\partial t\partial x}} + \frac{1}{m}\frac{{\partial S(x,t)}}{{\partial x}}\frac{{{\partial ^2}S(x,t)}}{{\partial {x^2}}} = \frac{{e\partial \phi (x,t)}}{{\partial x}} - \frac{{\partial \mu (x)}}{{\partial x}} + \frac{{\partial B(x,t)}}{{\partial x}},\\
&B(x,t) = \frac{{\zeta {\hbar ^2}}}{{8m{n^2}(x,t)}}\left\{ {2n(x,t)\frac{{{\partial ^2}n(x,t)}}{{\partial {x^2}}} - {{\left[ {\frac{{\partial n(x,t)}}{{\partial x}}} \right]}^2}} \right\}.
\end{align}
\end{subequations}
\textbf{Now using $u(x,t)=(1/m)\partial S(x,t)/\partial x$, with little mathematical work, one arrives at the continuity and momentum equations in (\ref{gs}).} The chemical potential, $\mu(n,T)$ of the arbitrary degenerate electron gas is defined through the generalized parametric equation of state (EoS). The EoS for isothermal compression of electron gas with arbitrary degeneracy is then given in terms of the Fermi integrals
\begin{subequations}\label{eosr}
\begin{align}
&n(\eta ,T) = \frac{{{2^{7/2}}\pi {m^{3/2}}}}{{{h^3}}}{F_{1/2}}(\eta) =  - \frac{{{2^{5/2}}{{(\pi m{k_B}T)}^{3/2}}}}{{{h^3}}}{\rm{L}}{{\rm{i}}_{3/2}}[ - \exp (\eta )],\\
&P(\eta ,T) = \frac{{{2^{9/2}}\pi {m^{3/2}}}}{{3{h^3}}}{F_{3/2}}(\eta) =  - \frac{{{2^{5/2}}{{(\pi m{k_B}T)}^{3/2}}({k_B}T)}}{{{h^3}}}{\rm{L}}{{\rm{i}}_{5/2}}[ - \exp (\eta )].
\end{align}
\end{subequations}
Note that in current analysis we use the Thomas-Fermi assumption where the temperature is fixed in accordance with the Fermi-Dirac single-particle distribution \cite{dya1,dya2}.

The main purpose of current research is to use the model for metallic compounds with fully degenerate electrons and strongly coupled ions in static background lattice, i.e. jellium model. The degeneracy parameter is $\eta=T/T_F$ in which $T$ and $T_F$ are respectively the temperatures for electron gas and Fermi energy level. Therefore, the parametric regime $\eta<1$($\eta>1$) denotes the degenerate(nondegenrate) electron gas regime. For fully degenerate metals or metallic nanoparaticles, we have the extreme case of $\eta\ll 1$, or equivalently, $T_F\gg T$. For elemental metals at room temperature, $E_F$ amounts to about $10^4$K or slightly higher, depending on their electron concentrations. In this limit the chemical potential equals the Fermi energy and all electrons reside below the well defined Fermi level. On the other hand, ion dynamic properties are characterized via the ion coupling parameter, $\Gamma=Z e^2/d k_B T_i$, in which $d$ is the average inter-ion distances and $T_i$ is the ion temperature which is much lower compared to that of electron gas, $T_i\ll T_e$, due to the large mass ratio. A good criterion for the weak and strong coupling cases are $\Gamma\ll 1$ and $\Gamma \gg 1$, respectively \cite{ichimaru0}. Therefore, metallic compounds with $r_s=2-6$, where $r_s=d/r_B$ being the Bruckener parameter and $r_B$ the Bohr radius, are categorized as strongly coupled material with ions fixed in crystal lattice. For simplicity, in current model we ignore the effect of phonon excitations on plasmonic properties of the electron gas. Moreover, the electron fluid with $r_s\ll 10^{-2}$ is regarded as nonrelativistic electron gas, while semiconductors have the parameter value of $r_s>25$. Because for semiconductors the Fermi temperature is very close to that of the electrons, these material are regarded as partially degenerate electron gas. The electron degeneracy starts at electron number density of $n\simeq 10^{18}$cm$^{-3}$ where the free electron wavefunctions start to overlap. However, the electron density of doped semiconductors can be much lower than this value. In fully degenerate elemental metals the electron concentration is typically in the range ($10^{21}-10^{23}$)cm$^{-3}$ characterizing them as fully degenerate quasineutral electron gas.

Here we would like to give a brief comment on applicability of Schr\"{o}dinger-Poisson model to study fermion quasiparticle systems, which was first developed by Manfredi and Haas from quantum hydrodynamic theory and Wigner-Poisson system \cite{manfredi}. The quantum hydrodynamic equations are obtained from moments of the Wigner distribution function under liming assumptions. They used a first-order plane-wave approximation for N-body wavefunctions with amplitudes independent of quasiparticle orbital. Therefore, the only orbital dependent part comes from the phase function associated with each orbital. However, such assumption may put limitations on the applicability of the model to systems in thermal contact with surrounding environment, since, the density matrix which is used to study the temporal evolution becomes independent of plasmon orbital and only depends on spacial density variations. Bonitz et. al. \cite{bonitz0} have given an extensive discussion on different formulations and possible improvements of the quantum hydrodynamic model. It has been noted that, due to the mentioned limitations, quantum hydrodynamic model gives an expression for plasmon dispersion which differs from that of the random phase approximation (RPA) in the low-frequency static limit \cite{michta,akbhd,mold}. Therefore, a kinetic correction factor, $\zeta$, is applied to the quantum force term in order for the hydrodynamic theory to be consistent with RPA. In the proceeding analysis, we use the kinetic-corrected Schr\"{o}dinger-Poisson model in order to obtain statefunctions of pure quasiparticle orbital. The energy eigenvalues of these orbital are used to construct appropriate mixed states using quasiparticle orbital enumeration and determine thermodynamic quantities, exactly in a similar manner as is done in the free electron theory of solids \cite{ash}.

\section{Pure Quasiparticle States}

In order to study the system (\ref{sp}) in the linear limit, for simplicity we ignore minor interaction terms such as the electron exchange and correlations. However, we will account to antisymmetric wavefunctions later on in this analysis. Appropriately normalized time-independent half-space jellium model of quasineutral electron gas of arbitrary degeneracy, constitutes of a coupled system of second-order differential equations, the so-called pseudoforce system
\begin{subequations}\label{pf}
\begin{align}
&\zeta\frac{{d^2{\Psi(x)}}}{{d{x^2}}} + \Phi(x) + E \Psi(x) + E = 0,\\
&\frac{{d^2{\Phi(x)}}}{{d{x^2}}} - {\Psi}(x) = 0,
\end{align}
\end{subequations}
where we have used a variable separation condition, ${\cal N}(x,t)=\psi(x)\psi(t)$, for quasi-stationary collective excitations and applied the normalization scheme, $\Psi(x)=\psi(x)/\sqrt{n_0}$ and $\Phi(x)=e\phi(x)/E_p$ with $E_p=\hbar\sqrt{4\pi e^2 n_0/m}$ being the plasmon energy of the system. Also, $E=(\epsilon-\mu_0)/E_p$ with $\epsilon$ being the energy eigenvalue of the collective quasiparticle (plasmons) as defined through $\epsilon\psi(t)=\hbar\omega\psi(t)=i\sqrt{\zeta}\hbar d\psi(t)/dt$ with $\omega$ being the eigenfrequency of quasiparticle excitations. Note that the space and time variables are also normalized, respectively, to the plasmon length $1/k_p$ with $k_p=\sqrt{2mE_p}/\hbar$ being the plasmon wavenumber and $\hbar/E_p$. Also, the Thomas-Fermi assumption is used for the chemical potential in the fully degenerate limit, where, the chemical potential variations are compensated by the local electrostatic potential of the electron gas. The solution to psudoforce model, (\ref{pf}), may be written as
\begin{equation}\label{wf}
\left[ {\begin{array}{*{20}{c}}
{\Phi (x)}\\
{\Psi (x)}
\end{array}} \right] = \frac{{A\zeta }}{\alpha }\left\{ {\begin{array}{*{20}{c}}
{{\Psi _0} + k_2^2\left( {{\Phi _0} + E} \right)}&{ - \left[ {{\Psi _0} + k_1^2\left( {{\Phi _0} + E} \right)} \right]}\\
{ - \left[ {\left( {{\Phi _0} + E} \right)/\zeta  + k_1^2{\Psi _0}} \right]}&{\left( {{\Phi _0} + E} \right)/\zeta  + k_2^2{\Psi _0}}
\end{array}} \right\}\left[ {\begin{array}{*{20}{c}}
{\cos ({k_1}x)}\\
{\cos ({k_2}x)}
\end{array}} \right],
\end{equation}
in which $\Phi_0$ and $\Psi_0$ characterize the initial functional values for the homogenous system and $A$ is a normalizing factor, where, we assumed $d\Psi(x)/dx|_{x=0}=d\Phi(x)/dx|_{x=0}=0$, for simplicity. The characteristic excitation wavenumbers $k_1$ and $k_2$ are given as
\begin{equation}\label{ks}
{k_1} = \sqrt {\frac{{E  - \alpha }}{{2\zeta }}},\hspace{3mm}{k_2} = \sqrt {\frac{{E  + \alpha }}{{2\zeta }}},\hspace{3mm}\alpha  = \sqrt {{E^2} - 4\zeta }.
\end{equation}
\textbf{Note the complementarity-like relation between the wave- and particle-like excitation wavenumbers, $k_1 k_2=1$. The solution (\ref{wf}) leads to the matter-wave energy dispersion $E=(1+\zeta k^4)/k^2$ in which $E$ and $k$ are normalized to $E_p$ and plasmon wavenumber $k_p$, respectively. The solution (\ref{wf}) has the same characteristic as the proposed de Broglie's double solution in the pilot-wave theory \cite{bohm0,nikolic}, where a single electron is guided through by collective electrostatic interactions among other electrons. The kinetic correction parameter depends on normalized density and temperature parameters}
\begin{equation}\label{xin}
\zeta(\sigma,\theta)  = \frac{{{\rm{L}}{{\rm{i}}_{3/2}}\left[ { - \exp (\sigma /\theta )} \right]{\rm{L}}{{\rm{i}}_{ - 1/2}}\left[ { - \exp (\sigma /\theta )} \right]}}{3{{\rm{L}}{{\rm{i}}_{1/2}}{{\left[ { - \exp (\sigma /\theta )} \right]}^2}}},
\end{equation}
in which $\sigma=\mu_0/E_p$, $\theta=T/T_p$ are fractional chemical potential and electron temperature parameters with the plasmon temperature defined as, $T_p=E_p/k_B$.

\section{Damped Quasiparticle States}

The quasipaticle excitations in (\ref{pf}) is now generalized to include the spacial damping or screening effect at the jellium boundary, $x=0$. To this end, we consider the damped pseudoforce system as follows \cite{akbnew}.
\begin{subequations}\label{dpf}
\begin{align}
&\zeta\frac{{{d^2}\Psi (x)}}{{d{x^2}}} + 2\zeta\xi\frac{{d\Psi (x)}}{{dx}} + \Phi (x) + {E}\Psi (x) +E = 0,\\
&\frac{{{d^2}\Phi (x)}}{{d{x^2}}} + 2\xi\frac{{d\Phi (x)}}{{dx}} - \Psi (x) = 0,
\end{align}
\end{subequations}
where $\xi=k_{sc}/k_p$ is the normalized one-dimensional screening parameter depending on the fractional parameters via, $\xi^2 = ({E_p}/2{n_0})\partial n/\partial \mu  = (1/2\theta ){\rm{L}}{{\rm{i}}_{1/2}}\left[ { - \exp ({\sigma}/\theta )} \right]/{\rm{L}}{{\rm{i}}_{3/2}}\left[ { - \exp ({\sigma}/\theta )} \right]$ \cite{akbnew}. The general solution to damped pseudoforce system (\ref{dpf}) is
\begin{subequations}\label{gd}
\begin{align}
\begin{array}{*{20}{l}}
{{\Phi _{d}}(x) = \frac{{A\zeta {{\rm{e}}^{ - \xi x}}}}{\alpha }\left\{ {\begin{array}{*{20}{l}}
{\left[ {k_2^2\left( {{\Phi _0} - E} \right) + {\Psi _0}} \right]\left[ {\cos ({\beta _1}x) + \frac{\xi }{{{\beta _1}}}\sin ({\beta _1}x)} \right] - }\\
{\left[ {k_1^2\left( {{\Phi _0} - E} \right) + {\Psi _0}} \right]\left[ {\cos ({\beta _2}x) + \frac{\xi }{{{\beta _2}}}\sin ({\beta _2}x)} \right]}
\end{array}} \right\},}\\
{{\Psi _{d}}(x) = \frac{{A\zeta {{\rm{e}}^{ - \xi x}}}}{\alpha }\left\{ {\begin{array}{*{20}{l}}
{\left[ {\left( {{\Phi _0} - E} \right)/\zeta  + k_2^2{\Psi _0}} \right]\left[ {\cos ({\beta _2}x) + \frac{\xi }{{{\beta _2}}}\sin ({\beta _2}x)} \right] - }\\
{\left[ {\left( {{\Phi _0} - E} \right)/\zeta  + k_1^2{\Psi _0}} \right]\left[ {\cos ({\beta _1}x) + \frac{\xi }{{{\beta _1}}}\sin ({\beta _1}x)} \right]}
\end{array}} \right\},}
\end{array}
\end{align}
\end{subequations}
where $\beta_1=\sqrt{k_1^2-\xi^2}$ and $\beta_2=\sqrt{k_2^2-\xi^2}$. The boundary values $\Phi_0$ and $\Psi_0$ are defined using the half-space electron density distribution condition. \textbf{For the spatial region, $x\le 0$, we use the solution (\ref{wf}), while, for $x>0$ the solution (\ref{gd}) is used. Let us assume $\Phi(x=0)=0$ at the boundary which leads to $\Phi_0=-E$. On the other hand, we assume that the electron gas is extended to $-L$, thus, leading to equation}
\begin{equation}\label{L}
\int\limits_{ - L}^0 {A|\Psi (x){|^2}dx = L},
\end{equation}
\textbf{in which $A$ is a dimensionless normalization factor for both $\Psi(x)$ and $\Phi(x)$. The condition (\ref{L}) gives rise to the following expression in terms of $A$ and $\Psi_0$}
\begin{subequations}\label{nf}
\begin{align}
\frac{{4k_1^2k_2^3\cos \left( {{k_1}L} \right)\sin \left( {{k_2}L} \right)}}{{L\left( {k_1^2 - k_2^2} \right)\left( {k_1^4 + k_2^4} \right)}} - \frac{{4k_1^3k_2^2\cos \left( {{k_2}L} \right)\sin \left( {{k_1}L} \right)}}{{L\left( {k_1^2 - k_2^2} \right)\left( {k_1^4 + k_2^4} \right)}} + \frac{{\left( {k_1^3 + k_2^3} \right)\sin \left( {2{k_1}L} \right)}}{{2L\left( {k_1^4 + k_2^4} \right)}} = \frac{{2{\alpha ^2}{{\left( {k_1^4 + k_2^4} \right)}^{ - 1}}}}{{{A^2}{\zeta ^2}\Psi _0^2}}.
\end{align}
\end{subequations}
\textbf{In the limit large system size as compared to plasmon length, $L\gg 1$, we arrive at a simplified expression}
\begin{equation}\label{psi0}
A(E,\sigma ,\theta ) = \frac{{\alpha \sqrt 2 }}{{\zeta (\sigma ,\theta ){\Psi _0}\sqrt {k_1^4 + k_2^4} }}.
\end{equation}
\textbf{We have taken $L=100$ throughout this analysis. It is found that the solution for the undamped region, $x<0$, reduces to that of the electron gas confined to infinite potential \cite{akbquant}}
\begin{equation}\label{wf0}
\left[ {\begin{array}{*{20}{c}}
{{\Phi}(x)}\\
{{\Psi}(x)}
\end{array}} \right] = \sqrt {\frac{2}{{k_1^4 + k_2^4}}} \left[ {\begin{array}{*{20}{c}}
1&{ - 1}\\
{ - k_1^2}&{k_2^2}
\end{array}} \right]\left[ {\begin{array}{*{20}{c}}
{\cos ({k_1}x)}\\
{\cos ({k_2}x)}
\end{array}} \right].
\end{equation}
\textbf{The energy eigenvalues of the pure quasiparticle quantum states are quantized with the condition, $k_2\pm k_1=2\pi n$ (the quantum number $n$ not to be confused with electron number density) leading to the energy eigenvalues}
\begin{equation}\label{en}
{E_n} = \frac{{1 + \zeta {{\left( {n\pi  + \sqrt {1 + {n^2}{\pi ^2}} } \right)}^4}}}{{{{\left( {n\pi  + \sqrt {1 + {n^2}{\pi ^2}} } \right)}^2}}},
\end{equation}
\textbf{with the $n=0$ ground state level corresponding to $k_1=k_2$ is the quantum beating level which coincides with the quasiparticle conduction level minimum at $E_0=1+\zeta$.}

\textbf{In order to develop a theory for mixed quasiparticle quantum states we follow exactly the free electron analogy by numeration of the quasiparticle orbitals considering that the pure state wavefunctions are complete, orthonormal and antisymmetric (obtained via N-dimensional Slater determinant) due to the pauli exclusion acting on electrons. However, an important difference between electrons and collective fermion quasiparticles is that the later species do not have spin degeneracy like electrons, hence, each quasiparticle occupies a distinct energy level. This is implied by the fact that each quasiparticle orbital corresponds to a free electron (spin-up and spin-down) level \cite{akbquant}, so that, that metallic density is characterized by the resonant condition, $E_p=2E_F$ \cite{akbground}. In other words, the change in single-electron spin in antisymmetric N-electron wavefunction modifies the corresponding quasiparticle orbital to a new level.}

\section{Mixed Quasiparticle States}

\textbf{Once the pure quasiparticle states are known, it is an easy task to compose appropriate mixed states for equilibrium condition. These states are given as}
\begin{subequations}\label{ms}
\begin{align}
&\Phi (x) = \sum\limits_j {f({E_j}){\Phi _j}(x)},\hspace{3mm}\Psi (x) = \sum\limits_j {f({E_j}){\Psi _j}(x)},\\
&n(x) = \sum\limits_j {{n_j}},\hspace{3mm}{n_j} = {g_j}|{\Psi _j}{|^2},
\end{align}
\end{subequations}
\textbf{where $f(E_j)$ is the probability of occupying the $j$-th orbital by a quasiparticle and $g_i=f(E_j)D(E_j)/\sum\limits_j f(E_j)D(E_j)$ ($\sum_j g_j=1$), with $D(E_j)$ being the density of states at given energy eigenvalue, $E_j$, is the occupation number at each orbital. Note that $n_j$ gives the number density of quasiparticles at given orbital which is exactly equal to the number of electrons. Note also that the mixed state-vector, $\{\Phi(x),\Psi(x)\}$, satisfies the corresponding pseudoforce systems (\ref{pf}) and (\ref{dpf}). This can be shown by appropriate weighted summation over orbital energy eigenvalues.}

\textbf{Let us now study the system by enumeration of quasiparticle modes. For $N$ electron orbital there corresponds $N$ quuasiparticle orbital. The quasiparticle wavenumber modes are given by $k = (N\pi  + \sqrt {1 + {N^2}{\pi ^2}} )/L$, corresponding to maximum orbital energy of}
\begin{equation}\label{em}
{E_m} = \frac{{{L^4} + \zeta {{\left( {N\pi  + \sqrt {1 + {N^2}{\pi ^2}} } \right)}^4}}}{{{L^2}{{\left( {N\pi  + \sqrt {1 + {N^2}{\pi ^2}} } \right)}^2}}}.
\end{equation}
\textbf{The quasiparticle DoS is $dN/dE$ which follows}
\begin{equation}\label{dos}
D(E) = \frac{{\left( {{E^2} - 4\zeta } \right)\left( {{L^4} + \zeta } \right) + E\left( {{L^4} - \zeta } \right)\sqrt {{E^2} - 4\zeta } }}{{4\pi L\sqrt 2 \left( {{E^2} - 4\zeta } \right)\sqrt {\zeta \left( {{L^4} - \zeta } \right)\sqrt {{E^2} - 4\zeta }  - 4{L^2}{\zeta ^2} + E\left( {{L^4} + \zeta } \right)} }}.
\end{equation}
\textbf{Therefore, we have $g_j=f(E_j)D(E_j)/\sum\limits_j f(E_j)D(E_j)$ where $f(E_j)=1/[1+\exp(E_j/\theta)]$ is the Fermi occupation number which we assume also applies to quasiparticles for reasons mentioned earlier. However, in the continuum limit for a large spatial system size compared to the plasmon length, $L\gg 1$, the electron number density distribution is given by}
\begin{equation}\label{density}
n(x) = \int\limits_{2\sqrt \zeta  }^\infty  {D(E)f(E){{\left| {\Psi_E (x)} \right|}^2}dE},
\end{equation}
\textbf{and the electrostatic energy distribution can be obtained also as follows}
\begin{equation}\label{potential}
\Phi(x) = \int\limits_{2\sqrt \zeta}^\infty  f(E)\Phi_E (x)dE.
\end{equation}

\section{Numerical Analysis and Discussion}

\begin{figure}[ptb]\label{Figure1}
\includegraphics[scale=0.71]{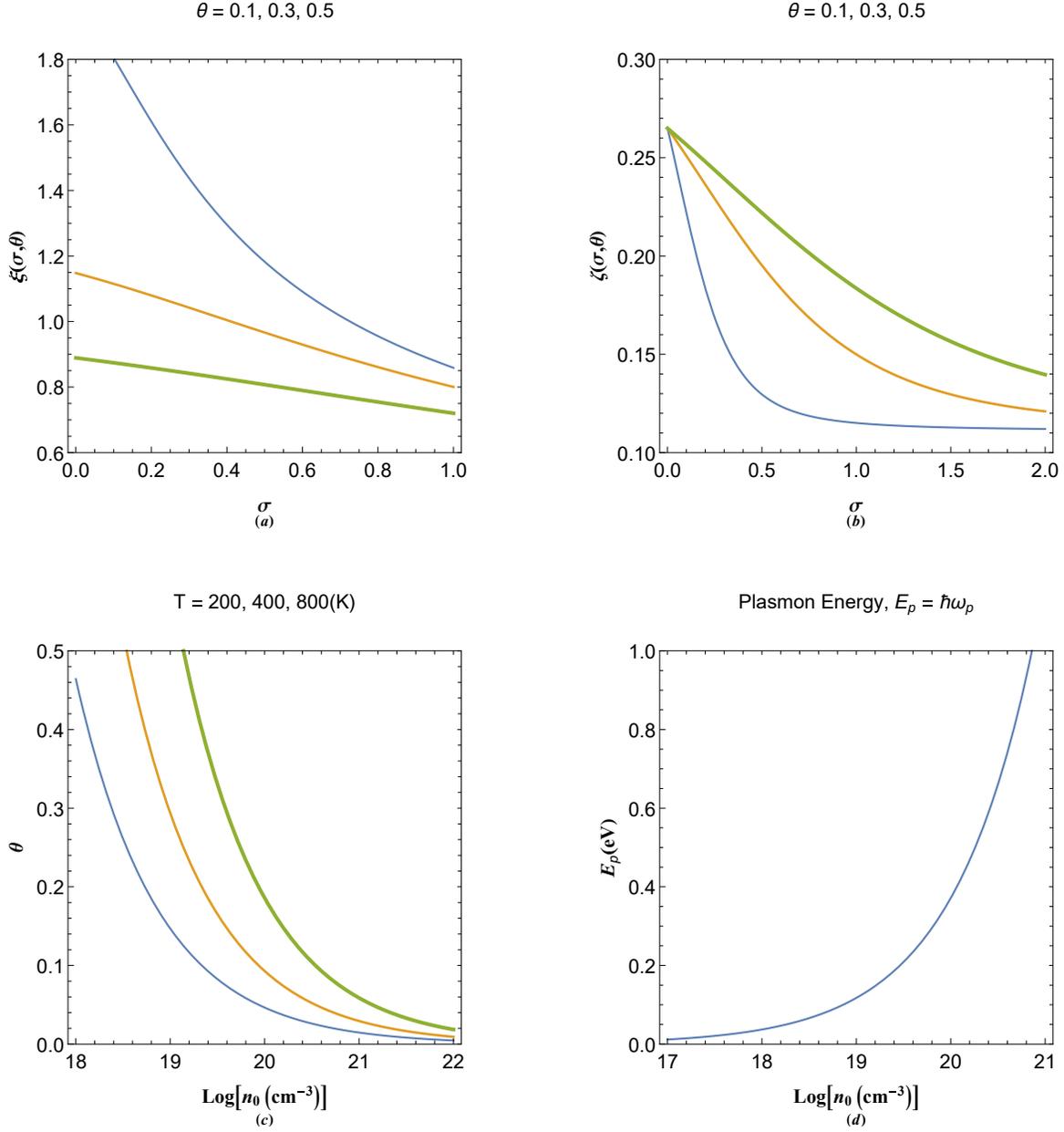}\caption{1(a) The variation of normalized screening (damping) parameter as a function of normalized chemical potential, $\sigma$ for different values of normalized electron temperature, $\theta$. 1(b) The variation of kinetic correction as a function of normalized chemical potential, $\sigma$ for different values of normalized electron temperature, $\theta$. 1(c) Variation in normalized temperature $\theta$ in terms of electron number density in logarithmic scale. 1(d) The plasmon energy variation in terms of the electron number density in logarithmic scale. The increase in the thickness of curves in each plot is meant to represent an increase in the varied parameters above each panel.}
\end{figure}

Figure 1 shows variations of the characteristic plasmon parameters for the electron gas of arbitrary degeneracy. Figure 1(a) depicts the variations of normalized one-dimensional screening parameter, $\xi(\mu,\theta)$ in terms of the normalized chemical potential, $\mu$, for various values of the normalized plasmon temperature, $\theta$. It is remarked that with increase in the chemical potential of the gas the screening parameter decreases substantially. It is also seen that as the normalized temperature increases, variations in $\xi$ over the chemical potential range decreases. This plot also shows that for the case of complete degeneracy the dependence of the normalized screening parameter to $\theta$ becomes insignificant. Figure 1(b) shows the dependence of kinetic correction parameter on the normalized chemical potential, $\sigma$, for different values of the normalized plasmon temperature, $\theta$. It is remarked that the correction decreases with decrease in $\sigma$. However, for given value of $\sigma>0$ the correction higher for larger value of $\theta$. The value of this parameter is always in the range $1/9<\zeta<0.265$. Figure 1(c) depicts the variations in the fractional electron temperature, $\theta$, as a function of the electron number density for various absolute electron temperatures. It is seen that for a given electron temperature the normalized electron temperature decreases significantly with increase in the electron number density. The variation over density becomes even more significant when the electron temperature increases. It is remarked that in the complete degeneracy limit dependence of this parameter to electron temperature becomes insignificant. Moreover, Fig. 1(d) shows the variation in plasmon energy in terms of electron number density in a logarithmic scale. It is remarked that the increase in plasmon energy becomes sharp as the degeneracy limit, ($n_0\simeq 10^{18}$cm$^{-3})$, sets in. For typical metals the plasmon energy amounts a few electron Volts. For instance, for cesium with the plasmon energy as low as $E_p\simeq 2.9$eV it can reach much higher value for aluminium with $E_p\simeq 15$eV. However, in order to compare the screening effect in these the difference in their chemical potential which is their Fermi energy at zero-temperature limit ($E^{Cs}_F=1.59$eV, $E^{Al}_F=11.7$eV) must be taken into account.

\begin{figure}[ptb]\label{Figure2}
\includegraphics[scale=0.71]{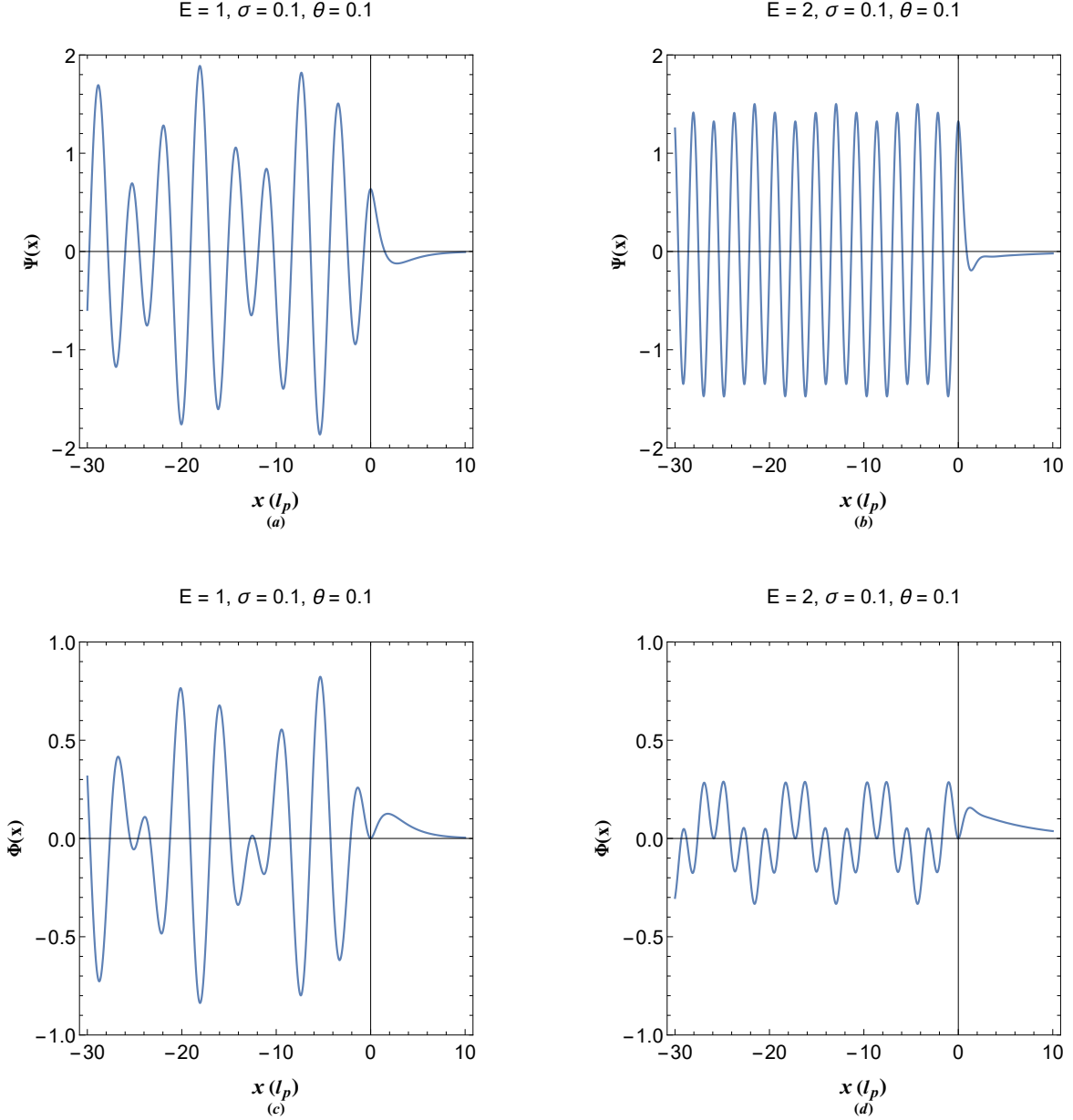}\caption{Profiles of pure quasiparticle state statefunctions of half-space electron gas of arbitrary degeneracy for given energy eigenvalue.}
\end{figure}

Figure 2 shows the variations in pure quasipaticle state functions, $\Psi(x)$, and electrostatic potential energy, $\Phi(x)$, for given values of normalized electron density and temperature and different energy eigenvalues. It is clearly evident that values and derivatives of state functions match at the half-space boundary. As remarked in Figs. 2(a) and 2(b), for $x<0$ there are dual-tone oscillations in quasiparticle probability, $\Psi(x)$ due to dual nature of these entities. Meanwhile, in the screened region, $x>0$, there is an oscillatory exponential decrease in amplitude of the quasiparticle state, $\Psi(x)$. In Fig. 2(b) the increase in energy eigenvalue leads to a relative decrease in the wavefunction amplitude. Local variations in the electrostatic energy of the system is shown in Figs. 2(c) and 2(d). It seems that the variations in the region, $x<0$, for $\Phi(x)$ is also dual-tone as compared to that of $\Psi(x)$. This is due to the fact that plasmon excitations are of dual-wavenumber character with the larger wavenumber corresponding to the single particle oscillations and the smaller one to the collective oscillation in the electron gas. It is therefore seen that at high values of quasiparticle energy the modulated fine oscillations in electrostatic energy profile correspond to the single particle excitationsand the larger oscillatory pattern is due to the collective plasmonic oscillations in the electron gas. On the other hand, in the region, $x>0$, the electrostatic energy decay has nearly a monotonic exponential decay without oscillations. Figure 2(d) reveals that unlike for the case of number density, increase of energy eigenvalue also significantly lowers the amplitude of electrostatic energy profile. However, in the region, $x>0$, the decay rate for lower energy eigenvalue in Fig. 2(c) is slightly faster as compared to that of the high energy profile, shown in Fig. 2(d).

\begin{figure}[ptb]\label{Figure3}
\includegraphics[scale=0.71]{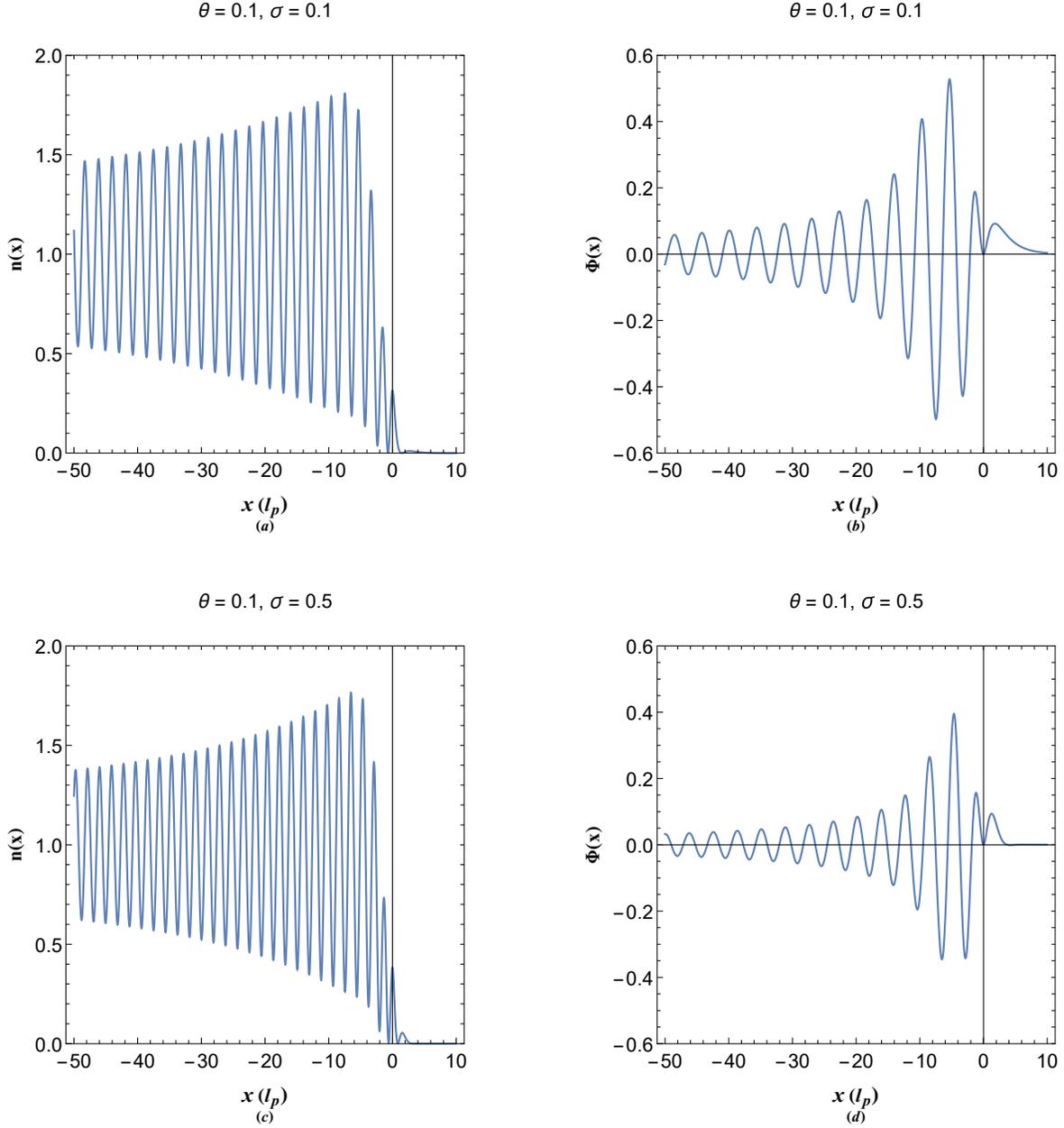}\caption{3(a) and 3(c) show the normalized perturbed electron number-density at a thermal equilibrium condition for given parameter values of $\theta$ and $\mu$. 3(b) and 3(d) show normalized perturbed electrostatic potential energy at a thermal equilibrium condition for given parameter values of $\theta$ and $\mu$.}
\end{figure}

Figure 3 depicts the spatial distribution of normalized perturbed electron number-density and electrostatic energy for given normalized chemical potential, $\sigma$, and temperature, $\theta$. Figure 3(a) shows the electron number density profile for given values of $\mu=0.3$ and $\theta=0.2$. The Fig. 3(a) illustrates some interesting features of half-space plasmon excitations at equilibrium temperature. It is seen that a density hump forms just before the jellium boundary, meanwhile, an electron depleted region beyond the jellium boundary appears just before a small electron density hump occurs. Another quite unique feature is the well-defined periodic density oscillations extended to the electron gas region ($x<0$) which is due to resonant interaction between single electron and collective excitations. Such feature has been shown to be a main characteristic feature of an electron gas confined to an infinite potential well \cite{akbquant}. The electron packet forming in front of the jellium boundary differs significantly form previous reported spill-out effect \cite{ciraci,mario2}. Figure 3(b) depicts the electrostatic energy profile corresponding to the parameter values used in Fig. 3(a). It is seen that the periodic structure is also present in this case. Close to the jellium boundary the amplitude of electrostatic energy is significantly increased, while, beyond the boundary the potential energy exponentially decreases towards infinity. Figures 3(c) and 3(d) depict the equilibrium mixed statefunction profiles for higher value of fractional chemical potential. It is evident from Fig. 3(c) that the density hump before physical boundary has become sharper. Note the formation of a surface dipole before the jellium boundary due to electron depletion befor the physical edge. It is furthe remarked that density of the electron packet beyond the physical edge increases significantly due to increase of the normalized chemical potential. Quite similar trend appears for the electrostatic energy profile as the value of $\sigma$ increases.

\begin{figure}[ptb]\label{Figure4}
\includegraphics[scale=0.71]{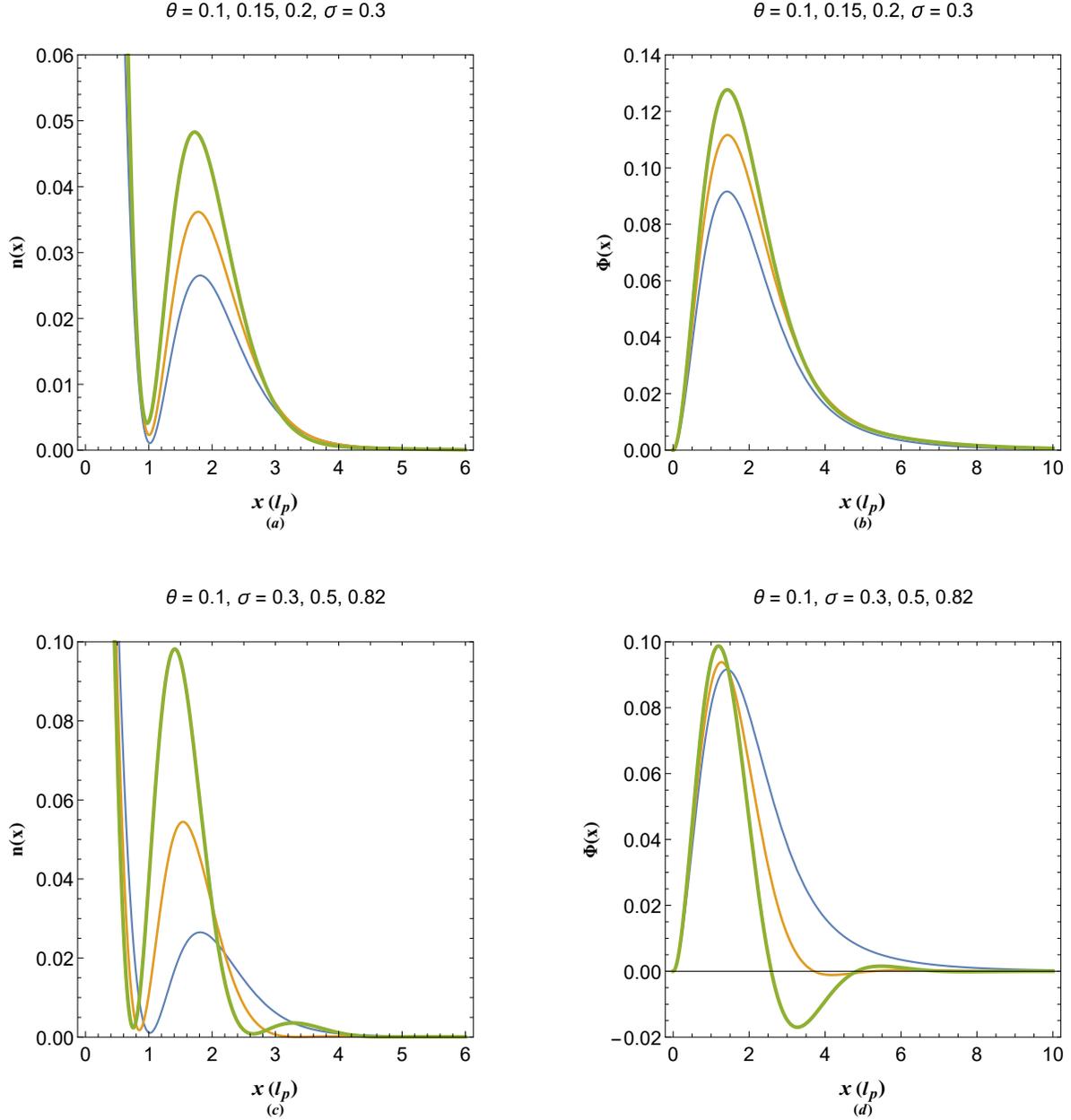}\caption{4(a) and 4(c) show the variations of normalized equilibrium electron number density and 4(b) and 4(d) show the normalized electrostatic potential energy beyond the jellium boundary for different values of the normalized electron temperature and chemical potential. Increase in thickness of curves in each plot represents an increase in the varied parameter above each panel.}
\end{figure}

Figure 4 depicts the variations of number density and electrostatic energy beyond the physical boundary, $x>0$. Figure 4(a) shows that the electron density extends to the screened area beyond the jellium boundary. Also, a there is density peak around $x=2$ which its amplitude increases with increase in the normalized electron temperature with the maximum value unchanged. Figure 4(b) shows the corresponding electrostatic energy indicating its maximum increase with the increase in the temperature. Figures 4(c) and 4(d) show the electron density and electrostatic energy profiles for different values of normalized chemical potential and fixed temperature. In Fig. 4(c) it is observed that the increase of chemical potential leads to sharp increase in electron density peak with its maximum value moved towards boundary. It is also remarked that Friedel-like oscillation appear for much higher chemical potential value characteristics of metallic density regime. The corresponding electrostatic energy profile is shown in Fig. 4(d) which reveals that increase in chemical potential leads to formation of oscillatory energy minimum in metallic chemical potential and temperature regime. The appearance of potential minimum beyond the physical surface of perfectly conducting metals is fully consistent with well-known Casimir-Polder effect \cite{niko1,niko2,bressi}. It can also describe the nanometallic particle agglomeration in quantum contact \cite{stefan}.

\begin{figure}[ptb]\label{Figure5}
\includegraphics[scale=0.71]{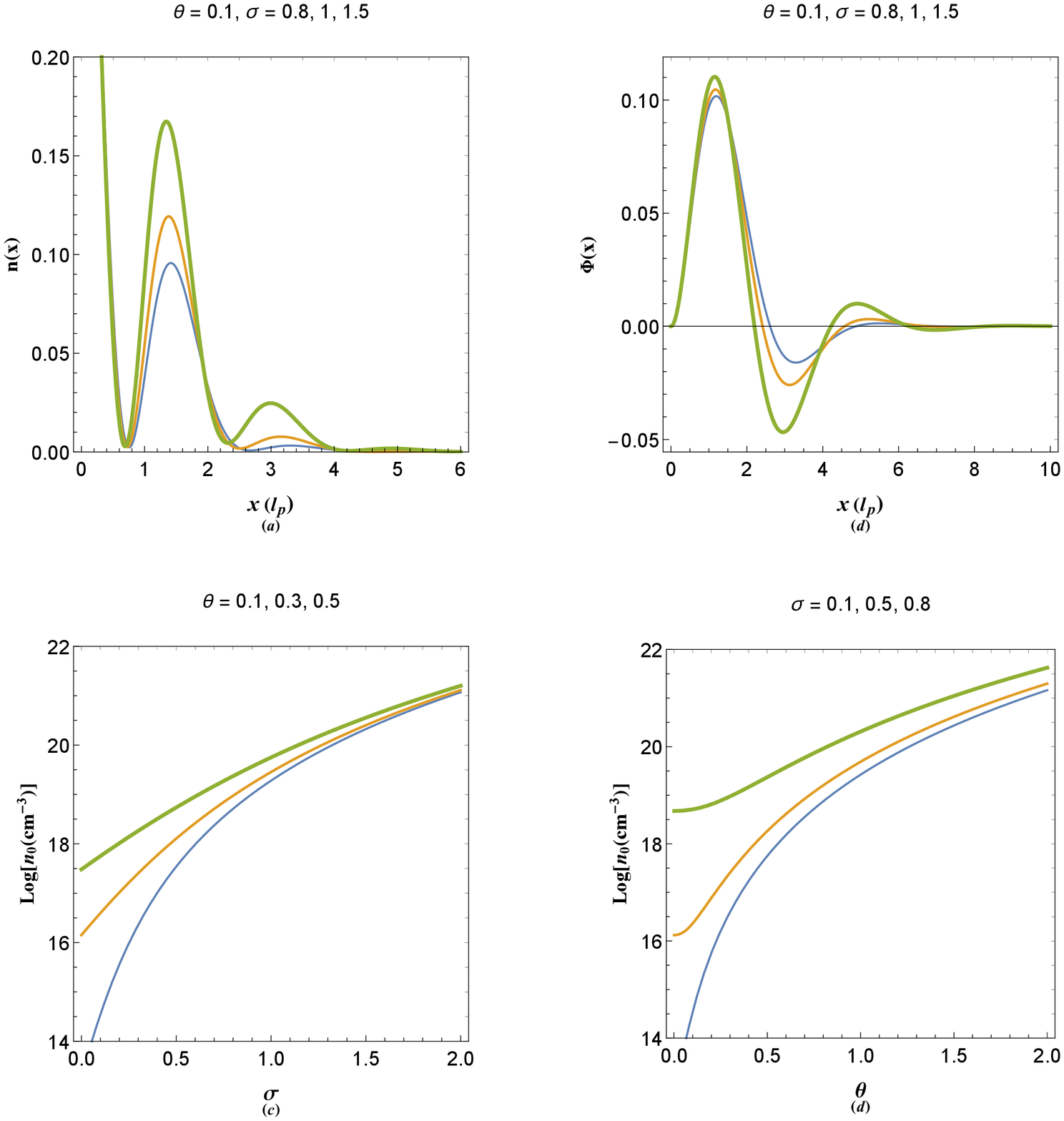}\caption{5(a) Variations in electrostatic energy profile in metallic chemical potential regime with the change in the normalized chemical potential of the electron gas. 5(b) Variations in electrostatic energy profile in metallic chemical potential regime with the change in the normalized electron temperature. 5(c) Variation of the electron number density with normalized chemical potential for different values of electron number density. 5(d) Variation of the normalized electron temperature with normalized chemical potential for different values of normalized electron temperature. The increase in the thickness of curves in each plot is meant to represent an increase in the varied parameter above each panel.}
\end{figure}

Figure 5 shows the variations in electron density and electrostatic potential profile with change in normalized temperature and chemical potential of the electron gas in dense metallic regime. Figure 5(a) reveals that increase of the chemical potential for fixed value of the electron temperature leads to sharp increase of the Friedel oscillation amplitude. Giant Friedel oscillations has been reported to exist in metallic surfaces such as beryllium \cite{spr}. On the other hand, Fig. 5(b) indicates that the increase of electron gas temperature in fixed chemical potential leads to increase in depth of the attractive potential energy, moving the potential minimum closer to the boundary. Based on our model, beryllium with a Fermi energy of $E_F\simeq 14.3$eV, as one of densest metallic elements, should exhibit a significant Casimir-Polder and nanoparticle agglomeration effects. Figures 5(c) and 5(d) show the variations of $\theta$ and $\mu$ with respect to the electron number density of electron gas. In terms of normalized parameters $\theta$ and $\sigma$ the electron number density can be written as ${n_0}(\sigma,\theta) = {n_p}{\theta ^6}{\rm{L}}{{\rm{i}}_{3/2}}{\left[ { - \exp (2{\sigma}/\theta )} \right]^4}$ with $n_p=16 e^6 m_e^3/(\pi^3 \hbar^6)\simeq 5.66\times 10^{19}$cm$^{-3}$ being a characteristic number-density. According to Fig. 5(c), the number density in $\sigma>1$ regime becomes almost independent of the value of $\theta$. This is reason for why the electron temperature plays the insignificant role in fully degenerate electron gas. It is therefore concluded that the oscillatory attractive potential energy may be present for electron densities of strongly doped N-type semiconductors. It is further confirmed from by Fig. 5(d) that for higher values of $\theta$, such as in the warm dense matter regime, the Friedel effect should also be present.

\begin{figure}[ptb]\label{Figure6}
\includegraphics[scale=0.73]{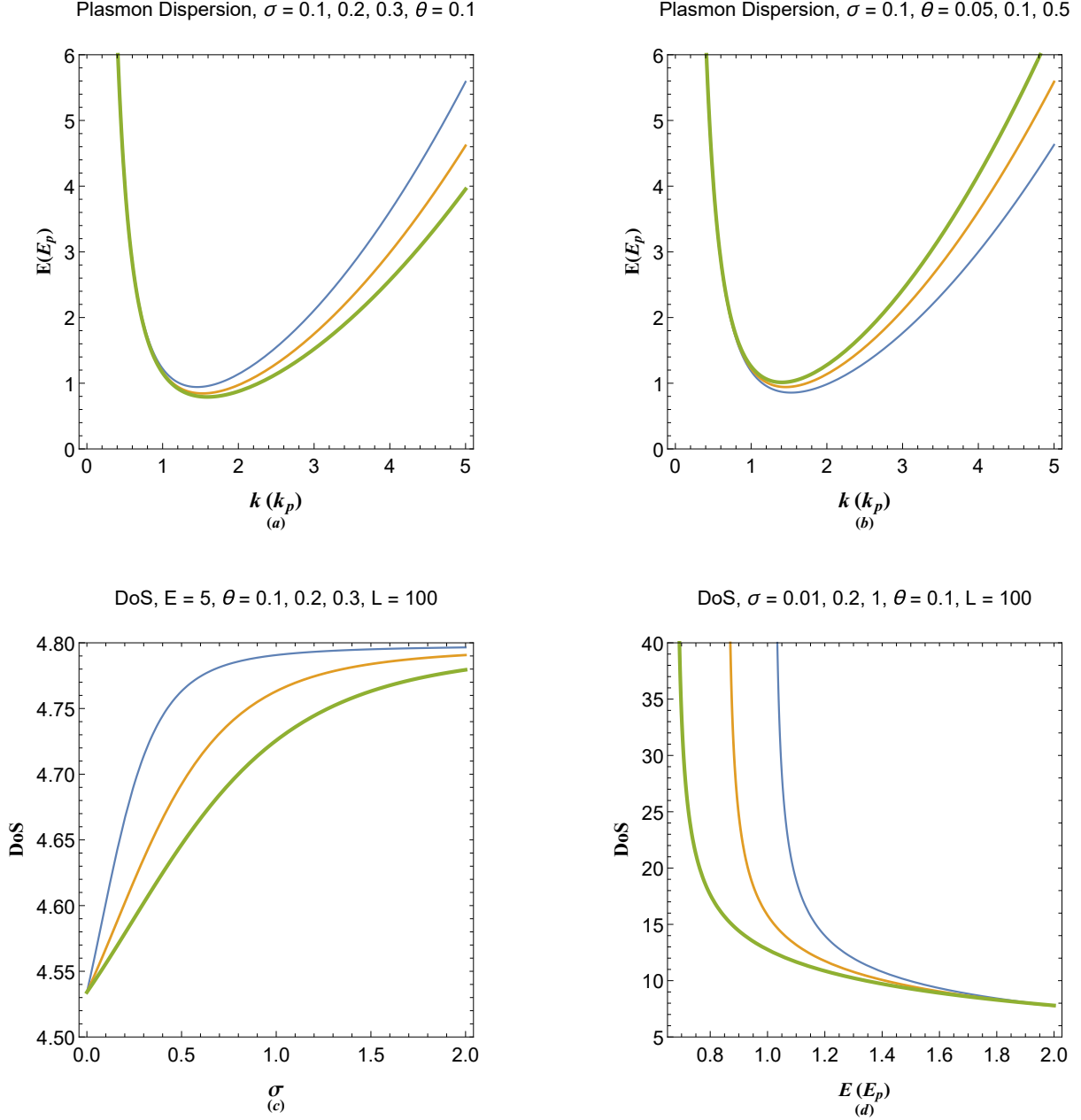}\caption{6(a) Variation in matter-wave dispersion of plasmon quasiparticle with changes in normalized chemical potential due to kinetic correction. 6(b) Variation in matter-wave dispersion of plasmon quasiparticle with changes in the normalized electron temperature due to kinetic correction. 6(c) Variation in the quasiparticle desity of states (DoS) with chemical potential and temperature. 6(d) Variation in the quasiparticle desity of states (DoS) with quasiparticle energy. The increase in the thickness of curves in each plot is meant to represent an increase in the varied parameter above each panel.}
\end{figure}

\textbf{Figure 6(a) shows variation in the energy dispersion of quasiparticle excitations with changes in chemical potential for fixed electron temperature. The plasmon quasiparticle conduction band minimum is at $E_m=2\sqrt{\zeta}$ and $k_m=\zeta^{-1/4}$. Therefore, it is seen that the variations come from the kinetic correction parameter, $\zeta(\sigma,\theta)$. It is revealed that the kinetic correction mostly affect the low phase speed particle-like branch of the dispersion. With increase in the chemical potential this branch moves to energy values. On the other hand, for fixed chemical potential, increase in the normalized electron temperature moves the low phase-speed branch to higher energy values. The curvature of the quasiparticle conduction band is related to the electron effective mass and consequently their mobility in collective motion. Figure 6(c) shows the variations in quasiparticle DoS with chemical potential and temperature of electron gas for given energy eigenvalue. It is remarked that the largest variation of DoS with the temeparture corresponds to the metallic regime $\sigma \simeq 1$. As the chemical potential increases, the DoS also increase for fixed temperature and reaches a maximum value for higher $\sigma$ values. However, increase of the temperature always lowers the quasiparticle DoS. Figure 6(d) shows the variations in DoS with energy eigenvalue. It is remarked that with increase in $E$ the DoS sharply decreases. This is contrary to the case of free electrons in which we have the relation $D_f(E)\propto \sqrt{E}$. It is also shown that increase in chemical potential significantly enhances the DoS at lower quasiparticle energies.}

\section{Conclusion}

In this research we used the pseudoforce model obtained from the kinetic-corrected quantum hydrodynamic theory in order to investigate the half-space fermion quasiparticle excitations in pure states. By using the appropriate boundary conditions and separation of variables the quantized energy eigenvalues of the pure states are obtained and appropriate mixed states were developed for thermal equilibrium state. The thermodynamic parameters such as the local electron density and electrostatic potential energy are calculated from the mixed quasiarticle states based on the density of states (DoS) and quasiparticle occupation probabilities. It was found that electron density packet develops in front of the free electron boundary profile of which is strongly dependent of the electron chemical potential and temperature. It was also revealed that for electron density and temperature range relevant to metallic regime Friedel like oscillations take place in the surface density distribution. Furthermore, attractive potential energy minimum appears in front of metallic surfaces which is consistent with the Casimir-Polder effect. The effect of quantum kinetic correction on the energy dispersion and DoS of fermion quasiparticle was also investigated in detail.

\end{document}